\documentclass[noshowpacs,amsmath,
twocolumn,
superscriptaddress,
8pt,
aps,prb]{revtex4-1}
\usepackage{setspace}
\usepackage{amsmath}
\usepackage{graphicx}
\usepackage[nearskip,margin = 0pt]{subfig}

\usepackage{verbatim}
\usepackage{amsfonts}
\usepackage{amssymb}
\usepackage{epstopdf} 
\usepackage{xcolor}
\usepackage{verbatim}
\usepackage{upgreek}
\usepackage{ragged2e}
\DeclareGraphicsExtensions{.pdf,.eps,.png,.jpg,.mps} 

\begin{document}
\title{Integrated optical frequency division for stable microwave and mmWave generation}

\author{Shuman Sun$^{1,*}$, Beichen Wang$^{1,*}$, Kaikai Liu$^{2}$, Mark Harrington$^{2}$, Fatemehsadat Tabatabaei$^{1}$, Ruxuan Liu$^{1}$, Jiawei Wang$^{2}$, Samin Hanifi$^{1}$, Jesse S. Morgan$^{1}$, Mandana Jahanbozorgi$^{1}$, Zijiao Yang$^{1,3}$, Steven Bowers$^{1}$, Paul Morton$^{4}$, Karl Nelson$^{5}$, Andreas Beling$^{1}$, Daniel Blumenthal$^{2}$ and Xu Yi$^{1,3,\dagger}$\\
\vspace{3pt}
$^1$Department of Electrical and Computer Engineering, University of Virginia, Charlottesville, Virginia 22904, USA.\\
$^2$Department of Electrical and Computer Engineering, University of California Santa Barbara, Santa Barbara, California 93016, USA.\\
$^3$Department of Physics, University of Virginia, Charlottesville, Virginia 22904, USA.\\
$^4$ Morton Photonics, Palm Bay, Florida 32905, USA.\\
$^5$ Honeywell International, Plymouth, Minnesota 55441, USA.\\
$^{\ast}$These authors contributed equally to this work.\\
$^{\dagger}$Corresponding author: yi@virginia.edu}

 
\date{\today}

\maketitle









\medskip


\noindent {\bf The generation of ultra-low noise microwave and mmWave in miniaturized, chip-based platforms can transform communication, radar, and sensing systems. Optical frequency division that leverages optical references and optical frequency combs has emerged as a powerful technique to generate microwaves with superior spectral purity than any other approaches. We demonstrate a miniaturized optical frequency division system that can potentially transfer the approach to a CMOS-compatible integrated photonic platform. Phase stability is provided by a large-mode-volume, planar-waveguide-based optical reference coil cavity and is divided down from optical to mmWave frequency by using soliton microcombs generated in a waveguide-coupled microresonator. Besides achieving record-low phase noise for integrated photonic microwave/mmWave oscillators, these devices can be heterogeneously integrated with semiconductor lasers, amplifiers, and photodiodes, holding the potential of large-volume, low-cost manufacturing for fundamental and mass-market applications.
}

\medskip

\noindent {\bf Introduction}

Microwave and mmWave with high spectral purity are critical for a wide range of applications, including metrology, navigation, and spectroscopy{\cite{essen1955atomic,long2015microwave,townes1975microwave}. Due to the superior fractional frequency stability of reference-cavity stabilized lasers when compared to electrical oscillators \cite{matei20171}, the most stable microwave sources are now achieved in optical systems by using optical frequency division \cite{fortier2011generation,xie2017photonic,nakamura2020coherent,li2023small} (OFD). Essential to the division process is an optical frequency comb \cite{cundiff2003colloquium,diddams2020optical}, which coherently transfers the fractional stability of stable references at optical frequencies to the comb repetition rate at radio frequency. The frequency division reduces the phase noise spectral density of the output microwave signal relative to the input optical signal by the square of the division ratio. A phase noise reduction greater than $10^9$ has been reported \cite{fortier2011generation}. However, so far, the most stable microwaves derived from optical frequency division rely on bulk or fiber-based optical references \cite{fortier2011generation,xie2017photonic,nakamura2020coherent,li2023small}, hindering the progress of applications in the field that demand exceedingly low microwave phase noise. 

\begin{figure*}[!bht]
\captionsetup{singlelinecheck=off, justification = RaggedRight}
\includegraphics[width=17cm]{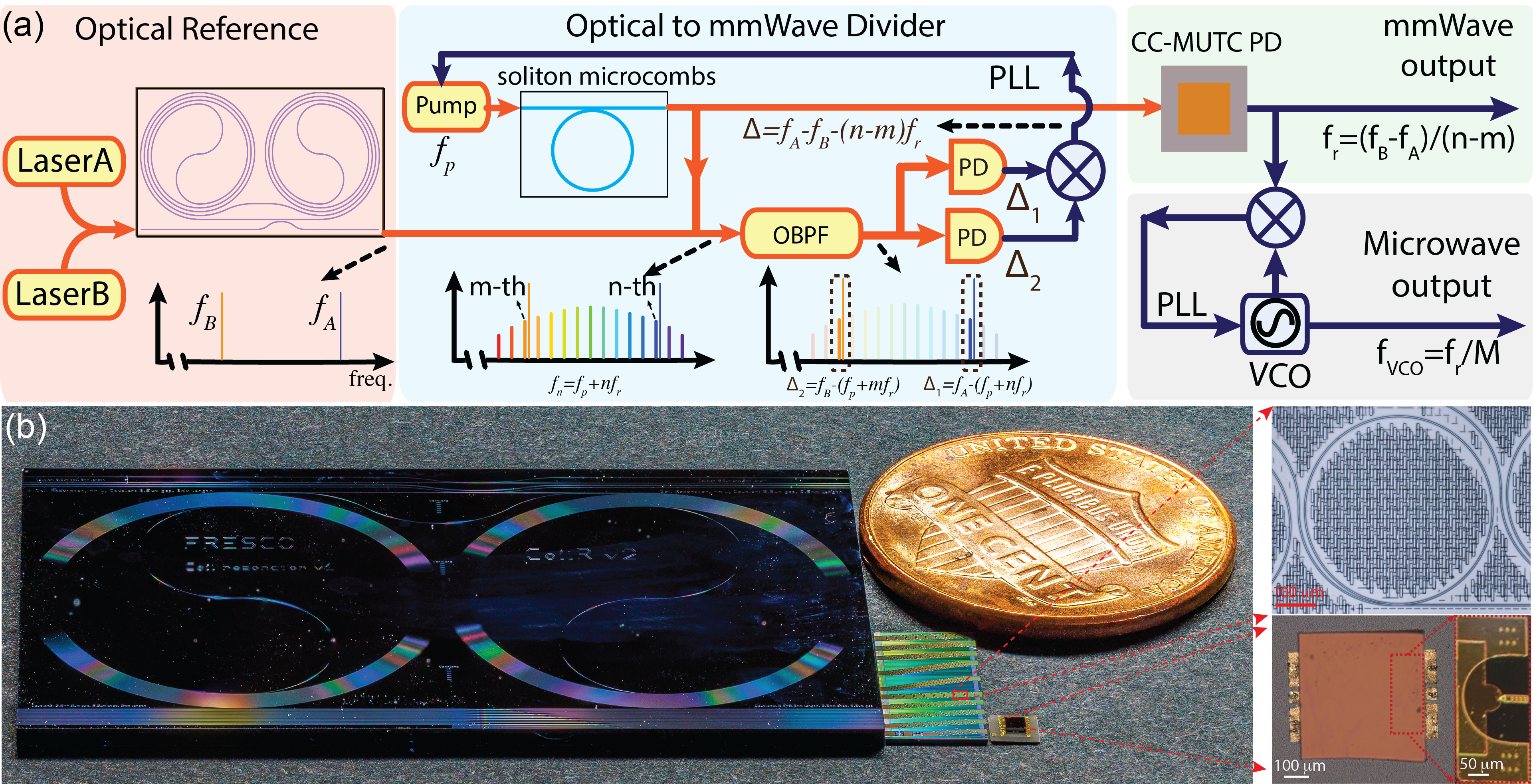}
\caption{{\bf Conceptual illustration of integrated optical frequency division.} {\bf (a)} Simplified schematic. A pair of lasers that are stabilized to an integrated coil reference cavity serve as the optical references and provide phase stability for the mmWave/microwave oscillator. The relative frequency difference of the two reference lasers are then divided down to the repetition rate of a soliton microcomb by feedback control of the frequency of the laser that pumps the soliton. High power, low noise mmWave is generated by photodetecting the OFD soliton microcomb on a CC-MUTC photodiode. The mmWave can be further divided down to microwave through a mmWave to microwave frequency division. {\bf (b)} Photo of critical elements in the integrated OFD. From left to right are: a SiN 4-meter long coil waveguide cavity as an optical reference, a SiN chip with tens of waveguide-coupled ring microresonators to generate soliton microcombs, a flip-chip bonded CC-MUTC PD for mmWave generation and a US one cent coin for size comparison. Microscopic pictures of a SiN ring resonator and a CC-MUTC PD are shown on the right.}
\label{fig:concept}
\end{figure*}

Integrated photonic microwave oscillators have been studied intensively for their potential of miniaturization and mass-volume fabrication. A variety of photonic approaches have been shown to generate stable microwave/mmWave signals, such as direct heterodyne detection of a pair of lasers \cite{kittlaus2021low}, microcavity-based stimulated Brillouin lasers \cite{li2013microwave,gundavarapu2019sub} and soliton microresonator-based frequency combs\cite{herr2014temporal,yi2017single,liu2020photonic,bai2021brillouin,yao2022soliton} (microcombs). For solid-state photonic oscillators, the fractional stability is ultimately limited by thermorefractive noise, which decreases with the increase of cavity mode volume \cite{gorodetsky2004fundamental,matsko2007whispering}. Large-mode-volume integrated cavities with meter-scale length and $>$ 100 million Q-factor have been shown recently \cite{lee2013spiral,puckett2021422} to reduce laser linewidth to Hz-level while maintaining chip footprint at centimeter-scale \cite{jin2021hertz,li2021reaching,liu202236}. However, increasing cavity mode volume reduces the effective intracavity nonlinearity strength and increases the turn-on power for Brillouin and Kerr parametric oscillation. This trade-off poses a difficult challenge for an integrated cavity to simultaneously achieve high stability and nonlinear oscillation for microwave generation. For oscillators integrated with photonic circuits, the best phase noise reported at 10 kHz offset frequency is demonstrated in the SiN photonic platform, reaching -109 dBc/Hz when the carrier frequency is scaled to 10 GHz \cite{liu2020photonic,jin2021hertz}. This is many orders of magnitude higher than that of the bulk OFD oscillators. An integrated photonic version of optical frequency division can fundamentally resolve this trade-off, as it allows the use of two distinct integrated resonators in OFD for different purposes: a large mode-volume resonator to provide exceptional fractional stability and a microresonator for the generation of soliton microcombs.  Together, they can provide major improvements to the stability of integrated oscillators.

Here, we significantly advance the state of the art in photonic microwave/mmWave oscillators by demonstrating integrated chip-scale optical frequency division. Our demonstration is based on CMOS-compatible SiN integrated photonic platform \cite{blumenthal2018silicon} and reaches record-low phase noise for integrated photonic-based microwave/mmWave oscillator systems. The oscillator derives its stability from a pair of commercial semiconductor lasers that are frequency stabilized to a planar waveguide-based reference cavity \cite{liu202236} (Fig. \ref{fig:concept}). The frequency difference of the two reference lasers is then divided down to mmWave with a two-point locking method \cite{li2014electro,tetsumoto2021optically} using an integrated soliton microcombs \cite{kippenberg2018dissipative,gaeta2019photonic}. While stabilizing soliton microcombs to long-fiber-based optical references has been shown very recently \cite{tetsumoto2021optically,kwon2022ultrastable}, its combination with integrated optical references has not been reported. The small dimension of microcavities allows soliton repetition rates to reach mmWave and THz frequencies \cite{zhang2019terahertz,tetsumoto2021optically,wang2021towards}, which have emerging applications in 5G/6G wireless communications \cite{koenig2013wireless,rappaport2019wireless}, radio astronomy \cite{clivati2017vlbi}, and radar \cite{ghelfi2014fully}.
Low noise, high power mmWaves are generated by photomixing the OFD soliton microcombs on a
high-speed flip-chip bonded charge-compensated modified uni-traveling carrier photodiode (CC-MUTC PD)\cite{xie2014improved,wang2021towards}. To address the challenge of phase noise characterization for high-frequency signals, a new mmWave to microwave frequency division (mmFD) method is developed to measure mmWave phase noise electrically while outputting a low-noise auxiliary microwave signal. The generated 100 GHz signal reaches a phase noise of -114 dBc/Hz at 10 kHz offset frequency (equivalent to -134 dBc/Hz for 10 GHz carrier frequency), which is more than two orders of magnitude better than previous SiN-based photonic microwave/mmWave oscillators \cite{liu2020photonic,jin2021hertz}. 
The ultra-low phase noise can be maintained while pushing the mmWave output power to 9 dBm (8 mW), which is only 1 dB below the record for photonic oscillators at 100 GHz \cite{sun2019high}. Pictures of chip-based reference cavity, soliton-generating microresonators, and CC-MUTC PD are shown in Fig. \ref{fig:concept}b.

\begin{figure*}[!bht]
\captionsetup{singlelinecheck=off, justification = RaggedRight}
\includegraphics[width=17cm]{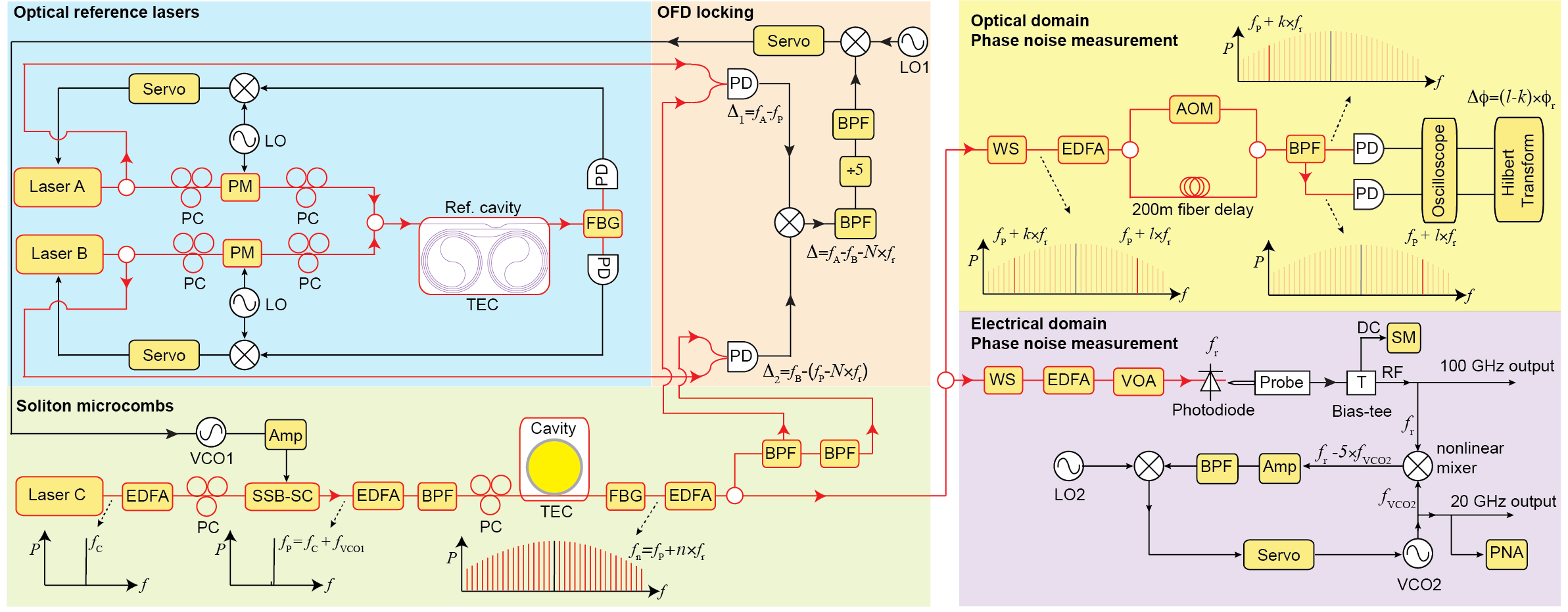}
\caption{{\bf Experimental setup.} A pair of reference lasers are created by stabilizing frequencies of lasers A and B to a SiN coil waveguide reference cavity, which is temperature controlled by thermoelectric cooler (TEC). Soliton microcomb is generated in an integrated SiN microresonator. The pump laser is the first modulation sideband of a modulated continuous wave (cw) laser, and the sideband frequency can be rapidly tuned by a voltage controlled oscillator (VCO). To implement two-point locking for optical frequency division, the 0-th comb line (pump laser) is photomixed with ref. laser A, while the $N$-th comb line is photomixed with ref. laser B. The two photocurrents are then subtracted on an electrical mixer to yield the phase difference between the ref. lasers and $N$ times the soliton repetition rate, which is then used to servo control the soliton repetition rate by controlling the frequency of the pump laser. The phase noise of the reference lasers and the soliton repetition rate can be measured in the optical domain by using dual-tone delayed self-heterodyne interferometry. Low noise, high power mmWaves are generated by detecting soliton microcombs on a CC-MUTC PD. To characterize the mmWave phase noise, a mmWave to microwave frequency division (mmFD) is implemented to stabilize a 20 GHz VCO to the 100 GHz mmWave, and the phase noise of the VCO can be directly measured by a phase noise analyzer (PNA). Erbium-doped fiber amplifiers (EDFAs), polarization controllers (PCs), phase modulators (PMs), single-sideband modulator (SSB-SC), band pass filters (BPFs), fiber Bragg grating (FBG) filters, line-by-line waveshaper (WS), acoustic-optics modulator (AOM), and electrical amplifiers (Amps) are also used in the experiment.}
\label{fig:setup}
\end{figure*}

\begin{figure*}[!bht]
\captionsetup{singlelinecheck=off, justification = RaggedRight}
\includegraphics[width=17cm]{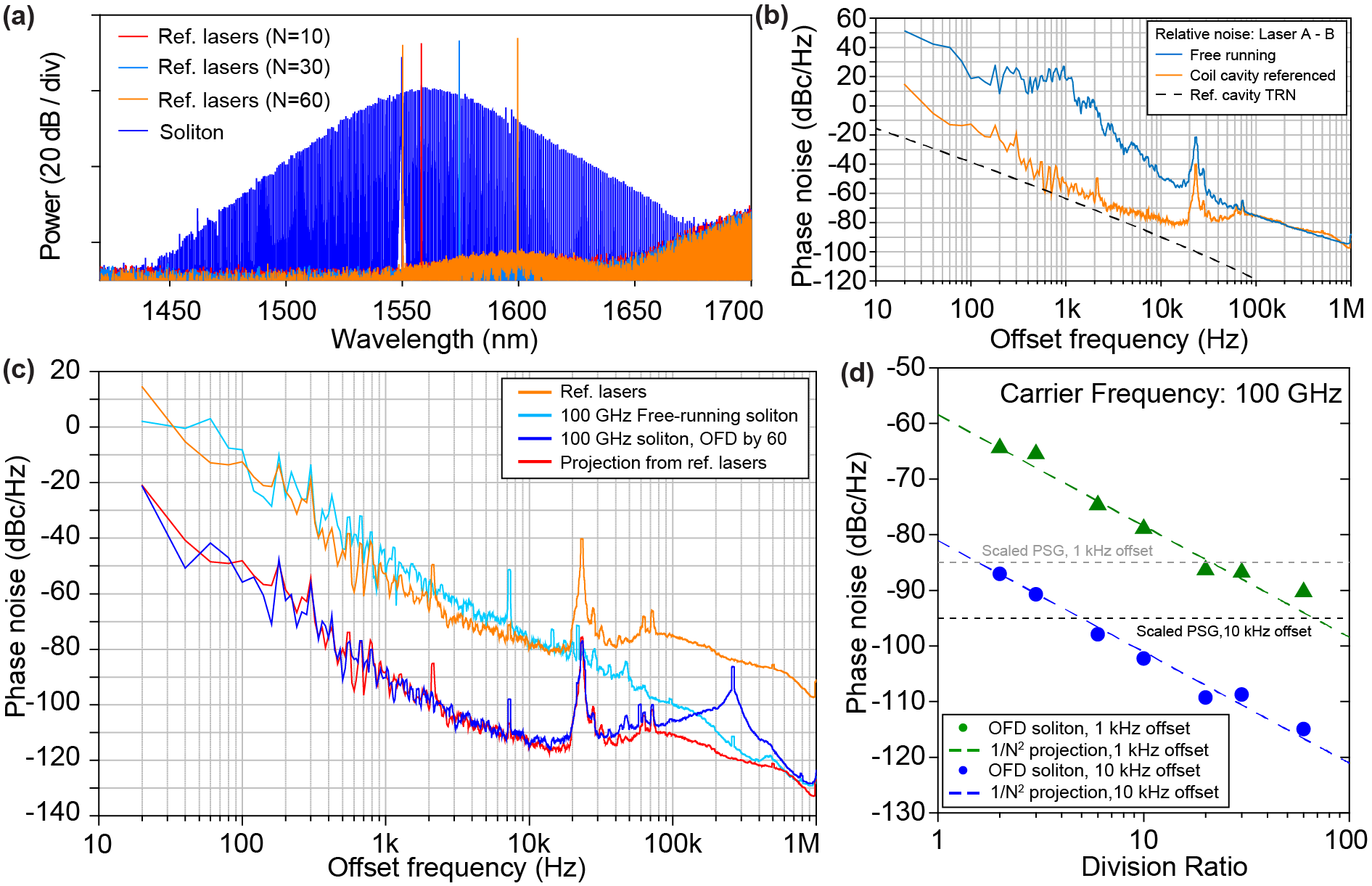}
\caption{{\bf Optical frequency division characterization.} {\bf (a)} Optical spectra of soliton microcombs (blue) and reference lasers corresponding to different division ratios. {\bf (b)} Phase noise of the frequency difference between the two reference lasers stabilized to coil cavity (orange) and the two lasers at free running (blue). The black dash line shows the thermal refractive noise (TRN) limit of the reference cavity. {\bf (c)} Phase noise of reference lasers (orange), the repetition rate of free-running soliton microcombs (light blue), soliton repetition rate after OFD with a division ratio of 60 (blue), and the projected repetition rate with 60 division ratio (red). {\bf (d)} Soliton repetition rate phase noise at 1 and 10 kHz offset frequencies versus OFD division ratio. The projections of OFD are shown with colored dashed lines.}
\label{fig:optical}
\end{figure*}

\begin{figure*}[!bht]
\captionsetup{singlelinecheck=off, justification = RaggedRight}
\includegraphics[width=17cm]{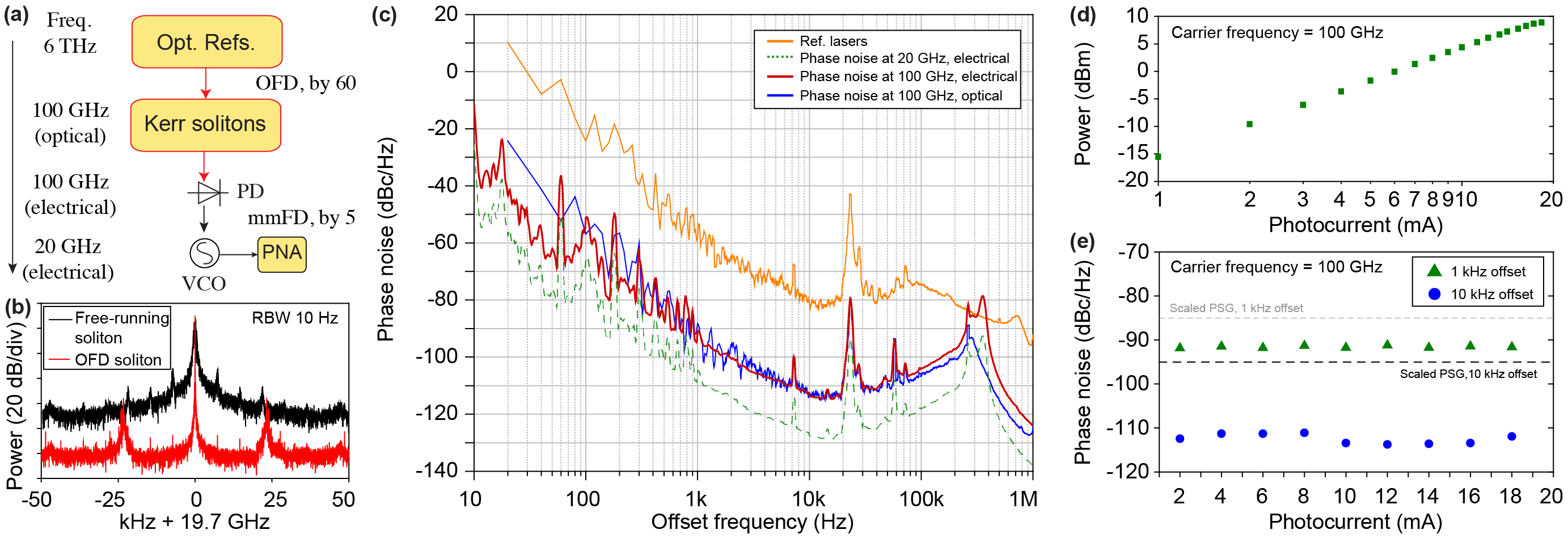}
\caption{{\bf Electrical domain characterization of mmWaves generated from integrated optical frequency division.} {\bf (a)} Simplified schematic of frequency division. The 100 GHz mmWave generated by integrated OFD is further divided down to 20 GHz for phase noise characterization. {\bf (b)} Typical electrical spectra of the VCO after mmWave to microwave frequency division (mmFD). The VCO is phase stabilized to the mmWave generated with OFD soliton (red) or free-running soliton (black). {\bf (c)} Phase noise measurement in the electrical domain. Phase noise of the VCO after mmFD is directly measured by the phase noise analyzer (dashed green). Scaling this trace to a carrier frequency of 100 GHz yields the phase noise upper bound of the 100 GHz mmWave (red). For comparison, phase noises of reference lasers (orange) and OFD soliton repetition rate (blue) measured in the optical domain are shown. {\bf (d)} Measured mmWave power versus PD photocurrent at -2 V bias. A maximum mmWave power of 9 dBm is recorded. {\bf (e)} Measured mmWave phase noise at 1 kHz and 10 kHz offset frequencies versus PD photocurrent.}
\label{fig:electrical}
\end{figure*}

\medskip

\noindent {\bf Results.}
The integrated optical reference in our demonstration is a thin-film SiN 4-meter-long coil cavity \cite{liu202236}. The cavity has a cross-section of 6 $\mu$m width $\times$ 80 nm height, a FSR of $\sim$ 50 MHz, an intrinsic quality factor of 41 $\times 10^{6}$ (41 $\times 10^{6}$) and a loaded quality factor of 34 $\times 10^{6}$ (31 $\times 10^{6}$) at 1550 nm (1599 nm). The coil cavity provides exceptional stability for reference lasers because of its large mode volume and high-quality factor \cite{}. Here, two widely tunable lasers (NewFocus Velocity TLB-6700, laser A and B) are frequency stabilized to the coil cavity through Pound–Drever–Hall (PDH) locking technique \cite{}. Their wavelengths can be tuned between 1550 nm ($f_A =193.4$ THz) and 1600 nm ($f_B =187.4$ THz), providing up to 6 THz frequency separation for optical frequency division. The setup schematic is shown in Fig.\ref{fig:setup}.

\medskip

The soliton microcomb is generated in an integrated, bus-waveguide coupled Si$_{3}$N$_{4}$ micro-ring resonator \cite{brasch2016photonic,wang2021towards} with a cross-section of 1.55 $\mu$m width $\times$ 0.8 $\mu$m height. The ring resonator has a radius of 228 $\mu$m, an FSR of 100 GHz, and an intrinsic (loaded) quality factor of 4.3 $\times 10^{6}$ (3.0 $\times 10^{6}$). The pump laser of the ring resonator is derived from the first modulation sideband of an ultra-low noise semiconductor extended distributed Bragg reflector (E-DBR) laser from Morton Photonics \cite{morton2018high}, and the sideband frequency can be rapidly tuned by a voltage-controlled oscillator (VCO). This allows single soliton generation by implementing rapid frequency sweeping of the pump laser\cite{stone2018thermal}, as well as fast servo control of the soliton repetition rate by tuning the VCO \cite{tetsumoto2021optically}. The optical spectrum of the soliton microcombs is shown in Fig. \ref{fig:optical}a, which has a 3-dB bandwidth of 4.6 THz. The spectra of reference lasers are also plotted in the same figure.

\medskip

The optical frequency division is implemented with the two-point locking method \cite{li2014electro,tetsumoto2021optically}. The two reference lasers are photomixed with the soliton microcomb on two separate photodiodes to create beat notes between the reference lasers and their nearest comb lines. The beat note frequencies are $\Delta_1 = f_A - (f_p + n\times f_r)$ and $\Delta_2 = f_B - (f_p + m\times f_r)$, where $f_r$ is the repetition rate of the soliton, $f_p$ is pump laser frequency, and $n,m$ are the comb line numbers relative to the pump line number. These two beat notes are then subtracted on an electrical mixer to yield the frequency and phase difference between the optical references and $N$ times of the repetition rate: $\Delta = \Delta_1 - \Delta_2 = (f_A - f_B) - (N\times f_r)$, where $N = n-m$ is the division ratio. Frequency $\Delta$ is then phase locked to a low-frequency local oscillator (LO, $f_\text{LO1}$) by feedback control of the VCO frequency. The tuning of VCO frequency directly tunes the pump laser frequency, which then tunes the soliton repetition rate through Raman self-frequency shift and dispersive wave recoil effects \cite{yi2016theory,yang2016spatial,yi2017single}. The frequency and phase of the optical references are thus divided down to the soliton repetition rate, as $f_r = (f_A - f_B - f_\text{LO1})/N$. Since the LO frequency is in the 10s MHz range and its phase noise is negligible compared to the optical references, the phase noise of the soliton repetition rate ($S_r$) within the servo locking bandwidth is determined by that of the optical references ($S_o$): $S_r = S_o/N^2$.

\medskip

To test the optical frequency division, the phase noise of OFD soliton repetition rate is measured for division ratios of $N=$ 2, 3, 6, 10, 20, 30, and 60. In the measurement, one reference laser is kept at 1550.1 nm, while the other reference laser is tuned to a wavelength that is $N$ microresonator FSR away from the first reference laser (Fig. \ref{fig:optical}a). The phase noise of the reference lasers and soliton microcombs are measured in the optical domain by using dual-tone delayed self-heterodyne interferometry \cite{kwon2017reference,jeong2020ultralow}. In this method, two lasers at different frequencies can be sent into an unbalanced Mach-Zehnder interferometer (MZI) with an acoustic optical modulator (AOM) in one arm (Fig. \ref{fig:setup}). Then the two lasers are separated by a fiber-Bragg grating (FBG) filter and detected on two different photodiodes. The instantaneous frequency/phase fluctuations of these two lasers can be extracted from the photodetector signals by using Hilbert transform. Using this method, the phase noise of the phase difference between the two stabilized reference lasers is measured and shown in Fig. \ref{fig:optical}b.
In this work, the phase noise of the reference lasers does not reach the thermal refractive noise limit of the reference cavity \cite{liu202236} and is likely to be limited by environmental acoustic and mechanical noise. For soliton repetition rate phase noise measurement, a pair of comb lines with comb numbers $l$ and $k$ are selected by a programmable line-by-line waveshaper (WS) and sent into the interferometry. The phase noise of their phase differences is measured, and its division by $(l-k)^2$ yields the soliton repetition rate phase noise \cite{jeong2020ultralow}.

\medskip

The phase noise measurement results are shown in Fig. \ref{fig:optical}c,d. The best phase noise for soliton repetition rate is achieved with a division ratio of 60 and is presented in Fig. \ref{fig:optical}c. For comparison,  the phase noises of reference lasers and the repetition rate of free-running soliton without OFD are also shown in the figure. Below 100 kHz offset frequency, the phase noise of OFD soliton is roughly $60^2 = 36$ dB below that of the reference lasers and matches very well with the projected phase noise for optical frequency division (noise of reference lasers - 36 dB). Phase noises at 1 kHz and 10 kHz offset frequencies are extracted for all division ratios and are plotted in Fig. \ref{fig:optical}d. The phase noises follow the $1/N^2$ rule, validating the optical frequency division. The measured phase noise for OFD soliton repetition rate is remarkably low for a mmWave oscillator. For comparison, 
phase noise of Keysight E8257D PSG signal generator (standard model) at 1 and 10 kHz are given in Fig. \ref{fig:optical}d after scaling the carrier frequency to 100 GHz. At 10 kHz offset frequency, our integrated OFD oscillator achieves a phase noise of -115 dBc/Hz, which is 20 dB better than a standard PSG signal generator. Interestingly, when comparing to integrated microcomb oscillators that are stabilized to long optical fibers \cite{tetsumoto2021optically}, our integrated oscillator matches the phase noise at 10 kHz offset frequency and provides better phase noise below 5 kHz offset frequency (carrier frequency scaled to 100 GHz). We speculate this is because our photonic chip is rigid and small when compared to fiber references, and thus are less affected by environmental noises like vibration and shock. This showcases the capability and potential of integrated photonic oscillators. 

\medskip

The OFD soliton microcomb is then sent to a high-power, high-speed flip-chip bonded charge-compensated modified uni-traveling carrier photodiode for mmWave generation. Similar to a uni-travelling carrier PD \cite{ishibashi1997high}, the carrier transport in the CC-MUTC PD depends primarily on fast electrons which provides high speed and reduces saturation effects due to space-charge screening. Power handling is further enhanced by flip-chip bonding the PD to a gold-plated coplanar waveguide on an aluminum nitride (AlN) submount for heat sinking \cite{beling2016high}. The PD used in this work is an 8-µm diameter CC-MUTC PD with 0.23 A/W responsivity at 1550 nm wavelength and a 3-dB bandwidth of 86 GHz. Details of the CC-MUTC PD are described elsewhere\cite{morgan2018high,li2016high}. While the power characterization of the generated mmWave is straightforward, phase noise measurement at 100 GHz is not trivial as the frequency exceeds the bandwidth of most phase noise analyzers (PNAs). One approach is to build two identical yet independent oscillators and down-mix the frequency for phase noise measurement. However, this is not feasible for us due to the limitation of lab resources. Instead, a new mmWave to microwave frequency division method is developed to coherently divide down the 100 GHz mmWave to 20 GHz microwave, which can then be directly measured on a phase noise analyzer (Fig. \ref{fig:electrical}a).

\medskip

In this mmWave to microwave frequency division, the generated 100 GHz mmWave and a 19.7 GHz VCO signal are sent to a highly nonlinear RF mixer, which creates higher harmonics of the VCO frequency to mix with the mmWave. The mixer outputs the frequency difference between the mmWave and the fifth harmonics of the VCO frequency: $\Delta f = f_r - 5f_\text{VCO2}$, and $\Delta f$ is set to be around 1.16 GHz. $\Delta f$ is then phase locked to a stable local oscillator ($f_\text{LO2}$) by feedback control of the VCO frequency. This stabilizes the frequency and phase of the VCO to that of the mmWave within the servo locking bandwidth, as $f_\text{VCO2} = (f_r - f_\text{LO2})/5$. The electrical spectrum and phase noise of the VCO are then measured directly on the phase noise analyzer and are presented in Fig. \ref{fig:electrical}b,c. The phase noise of the 19.7 GHz VCO can be scaled back to 100 GHz to represent the upper bound of the mmWave phase noise. For comparison, phase noise of reference lasers and OFD soliton repetition rate measured in the optical domain with dual-tone delayed self-heterodyne interferometry method are also plotted. Between 100 Hz to 100 kHz offset frequency, the phase noise of soliton rep-rate and the generated mmWave match very well with each other. This validates the mmFD method and indicates that the phase stability of the soliton rep-rate is well transferred to the mmWave. Below 100 Hz offset frequency, measurements in the optical domain suffer from phase drift in the 200-meter optical fiber in the interferometry and thus yield phase noise higher than that measured with the electrical method. Above 100 kHz offset frequency, the mmFD measured phase noise is limited by the strong mmFD servo locking bump. 


\medskip

Finally, the mmWave phase noise and power are measured versus the MUTC PD photocurrent from 1 to 18.3 mA at -2 Volt bias by varying the illuminating optical power on the PD. While the mmWave power increases with the photocurrent (Fig. \ref{fig:electrical}d), the phase noise of the mmWave remains almost the same for all different photocurrents (Fig. \ref{fig:electrical}e). This suggests that low phase noise and high power are simultaneously achieved. The achieved power of 9 dBm is one of the highest powers ever reported at 100 GHz frequency for photonic oscillators\cite{sun2019high}.

\medskip

\noindent {\bf Summary.}

\noindent In summary, we have demonstrated low noise mmWave and microwave generation by using integrated optical frequency division. The achieved phase noise is primarily limited by the technical noise in the reference lasers and can be improved in the future to thermorefractive noise (TRN) limit of the reference cavity by packaging the reference cavity to isolate environmental noises \cite{liu202236}. Common mode noise cancellation can be further leveraged at that point to reduce the relative noise between the two reference lasers to below the TRN limit. In addition, interesting developments in vacuum-gap cavities using microfabricated mirrors have been shown to overcome the TRN limit of planar waveguide reference cavities \cite{guo2022chip}. Chip-based stimulated Brillouin lasers are useful in reducing phase noise of optical references at high offset frequency \cite{lee2012chemically,gundavarapu2019sub}. The division ratio demonstrated in this work is limited by the frequency range of our tunable lasers instead of the optical span of the soliton microcombs. Microcomb-based OFD that leverages 10s THz optical span or even octave span is possible by using dispersion-engineered microresonators \cite{spencer2018optical,wu2023vernier}. Furthermore, recent development in integration of high-Q resonators and stress-optic modulator \cite{liu2020monolithic,wang2022silicon} will enable feedback control of microresonator frequency for OFD, which can greatly increase OFD servo bandwidth. Therefore, despite the fact that our demonstration has improved the phase noise of integrated photonic microwave/mmWave oscillators by more than 20 dB, it is feasible to further advance the phase noise by several orders of magnitude in the future, allowing integrated photonic oscillators to, in certain offset frequency range, e.g., 10 kHz, match some of the best bulk optical OFD oscillators \cite{fortier2011generation}. Finally, there is potential for fully integrated OFD oscillators through heterogeneous integration of SiN reference cavities and microresonators with other components \cite{xiang2022silicon}, e.g.,  semiconductor lasers, optical amplifiers, and photodiodes. The small size and mass of integrated oscillators will also offer improved performance regarding acceleration and shock.




\medskip





\medskip

\noindent Note: The authors would like to draw the readers’ attention to other integrated SiN optical frequency division work\cite{zhao2023all}, which was reported while preparing this manuscript.

\medskip

{\noindent \bf Data availability.} The data that support the plots within this paper and other findings of this study are available from the corresponding author upon reasonable request.

\medskip

\noindent\textbf{Acknowledgement}

\noindent The authors acknowledge Madison Woodson and Steven Estrella from Freedom Photonics for MUTC-PD fabrication, Ligentec for SiN microresonator fabrication, K. Vahala at Caltech for the access of tunable optical filters, and gratefully acknowledge DARPA GRYPHON (HR0011-22-2-0008), National Science Foundation (2023775), Advanced Research Projects Agency—Energy (DE-AR0001042). The views and conclusions contained in this document are those of the
authors and should not be interpreted as representing official policies of DARPA, ARPA-E, or the U.S. Government.

\bibliographystyle{naturemag}
\bibliography{ref}  

\setcounter{figure}{0}
\renewcommand{\figurename}{Extended Data FIG.}

\end{document}